# A Powder Diffraction-AI Solution for Crystalline Structure

Di Wu[1,2,3,#], Pengkun Wang[1,2,3,#], Shiming Zhou[3,4], Bochun Zhang[1,4], Liheng Yu[3], Xi Chen[3], Xu Wang[3], Zhengyang Zhou[3], Yang Wang[1,2,3,✉], Sujing Wang[1,3,4,5,✉], Jiangfeng Du[1,4,5]

[1]CAS Key Laboratory of Microscale Magnetic Resonance, University of Science and Technology of China, Hefei, China

[2]Key Laboratory of Precision and Intelligent Chemistry, University of Science and Technology of China, Hefei, China

[3]Suzhou Institute for Advanced Research, University of Science and Technology of China, Suzhou, China

[4]Hefei National Research Center for Physical Sciences at the Microscale, University of Science and Technology of China, Hefei, China

[5]Hefei National Laboratory, University of Science and Technology of China, Hefei, China

[#]These authors contributed equally: Di Wu and Pengkun Wang

[✉]*e*-mail: angyan@ustc.edu.cn (Y.W); sjwang4@ustc.edu.cn (S.W.)

## Abstract

Determining the atomic-level structure of crystalline solids is critically important across a wide array of scientific disciplines. The challenges associated with obtaining samples suitable for single-crystal diffraction, coupled with the limitations inherent in classical structure determination methods that primarily utilize powder diffraction for most polycrystalline materials, underscore an urgent need to develop alternative approaches for elucidating the structures of commonly encountered crystalline compounds. In this work, we present an artificial intelligence-directed leapfrog model capable of accurately determining the structures of both organic and inorganic-organic hybrid crystalline solids through direct analysis of powder X-ray diffraction data. This model not only offers a comprehensive solution that effectively circumvents issues related to insoluble challenges in conventional structure solution methodologies but also demonstrates applicability to crystal structures across all conceivable space groups. Furthermore, it exhibits notable compatibility with routine powder diffraction data typically generated by standard instruments, featuring rapid data collection and normal resolution levels.

The elucidation of crystalline solid structures with atomic-level precision is of paramount importance for the advancement of science and technology across a multitude of disciplines, including chemistry, physics, materials science, as well as medical and pharmaceutical research[1-3]. The crystal diffraction technique utilizing various radiation sources is widely recognized for its adaptability in achieving this objective. In recent decades, single-crystal diffraction techniques have demonstrated remarkable efficacy in analyzing single-crystal samples ranging from micrometers to nanometers. This evolution signifies a transition from traditional X-ray diffraction to advanced electron diffraction methodologies[4,5]. However, the intrinsic limitations associated with certain chemicals' propensity to yield single crystals suitable for structural determination have significantly restricted their practical applicability. As a result, the use of polycrystalline (powder) diffraction data has emerged as the predominant alternative for most crystalline solids. This approach typically requires an exemplary set of diffraction data derived from a combination of high-quality samples, sophisticated data collection techniques, and labor-intensive solution processes conducted by skilled crystallographers[6]. This requirement starkly contrasts with the prevailing reality characterized by numerous constraints—such as specimens exhibiting suboptimal crystallinity, limited access to state-of-the-art diffractometers, and researchers possessing minimal or even nonexistent expertise in structural elucidation.

To address this formidable challenge, a series of groundbreaking studies focused on the prediction and discovery of inorganic compounds have illuminated the notable capabilities and potential inherent in various artificial intelligence (AI) methodologies. These advancements have significantly enhanced the application of powder diffraction data in recent years[7,8]. Recently, a noteworthy breakthrough employing a deep-learning approach to tackle the crystallographic phase problem has begun to integrate organic and inorganic-organic hybrid solids into this rapidly evolving research domain[9]. This innovation offers a promising solution to one of the most pressing challenges in elucidating crystal structures from powder diffraction data, while simultaneously underscoring the persistent complexities associated with deriving precise structural models directly from such datasets. Following conventional logic, it is straightforward to systematically address each pivotal stage in resolving crystalline structures derived from traditional powder diffraction data—encompassing indexing, decomposition, solution determination, and refinement. However, this approach not only leads to a complex integration of stepwise intensive computational work but also encounters inevitable challenges that remain insurmountable due to fundamental principles[6]. Therefore, it becomes imperative to develop an innovative methodology capable of obtaining reliable crystal structures through direct analysis of powder diffraction data using unequivocal parameters while avoiding those associated with insoluble problems.

Here we present UstcUnfold, a leapfrog model capable of accurately determining the structures of both organic and inorganic-organic hybrid crystalline solids through direct analysis of powder X-ray diffraction (PXRD) data. A fundamental pattern-structure paired template library, derived from the Cambridge Crystallographic Data Centre (CCDC) database, has been developed to elucidate the overlapping high-dimensional information embedded within the input spectra. This advancement enables the generation of a preliminary structural model based on the input spectra and corresponding queried templates through a multi-task framework. Subsequently, it optimizes the predicted structure at an atomic level by employing an innovative chemically constrained residual diffusion network. Consequently, UstcUnfold provides an end-to-end solution that effectively circumvents parameters associated with insoluble challenges in conventional structure solution methodologies (Fig. 1). Furthermore, this model is applicable to crystal structures across all possible space groups, demonstrating unprecedented adaptability to diverse datasets. The most significant feature of this model is its notable compatibility with routine PXRD patterns commonly produced by standard instruments, allowing for rapid data collection and normal resolution levels.

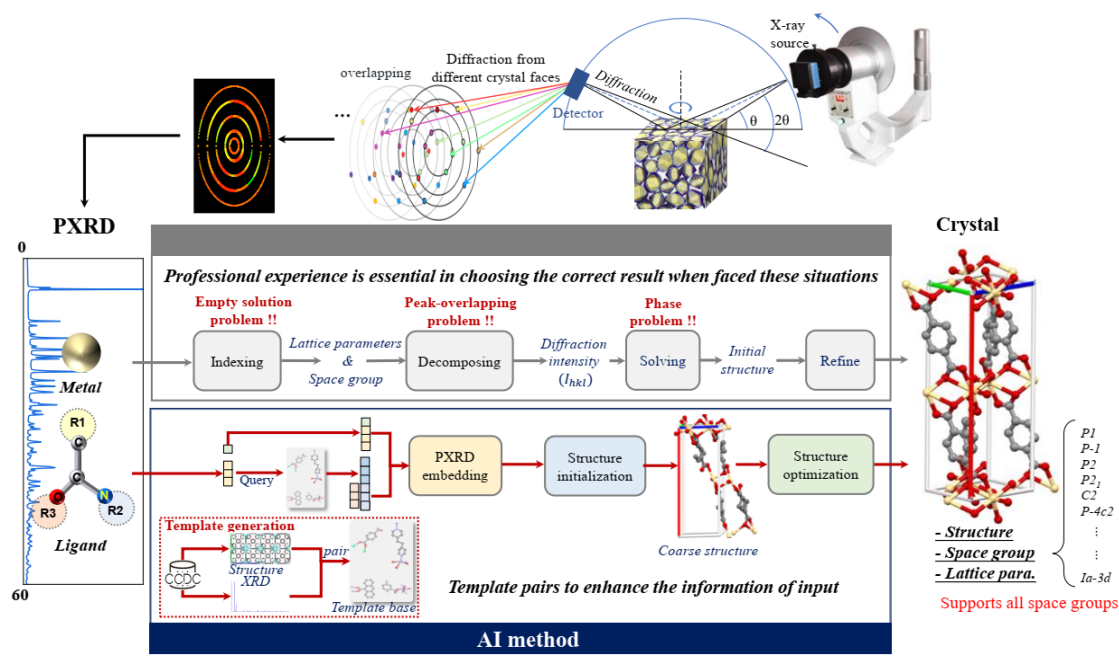

**Fig. 1│ UstcUnfold precisely determines organic and coordination crystal structures from PXRD data.** The upper section addresses the issue of overlapping PXRD peaks, along with the processes and challenges associated with conventional structure determination based on PXRD data. The lower section offers an overview of UstcUnfold, which encompasses template extraction and the use of a template library to enhance information for structural generation, followed by subsequent structure optimization.

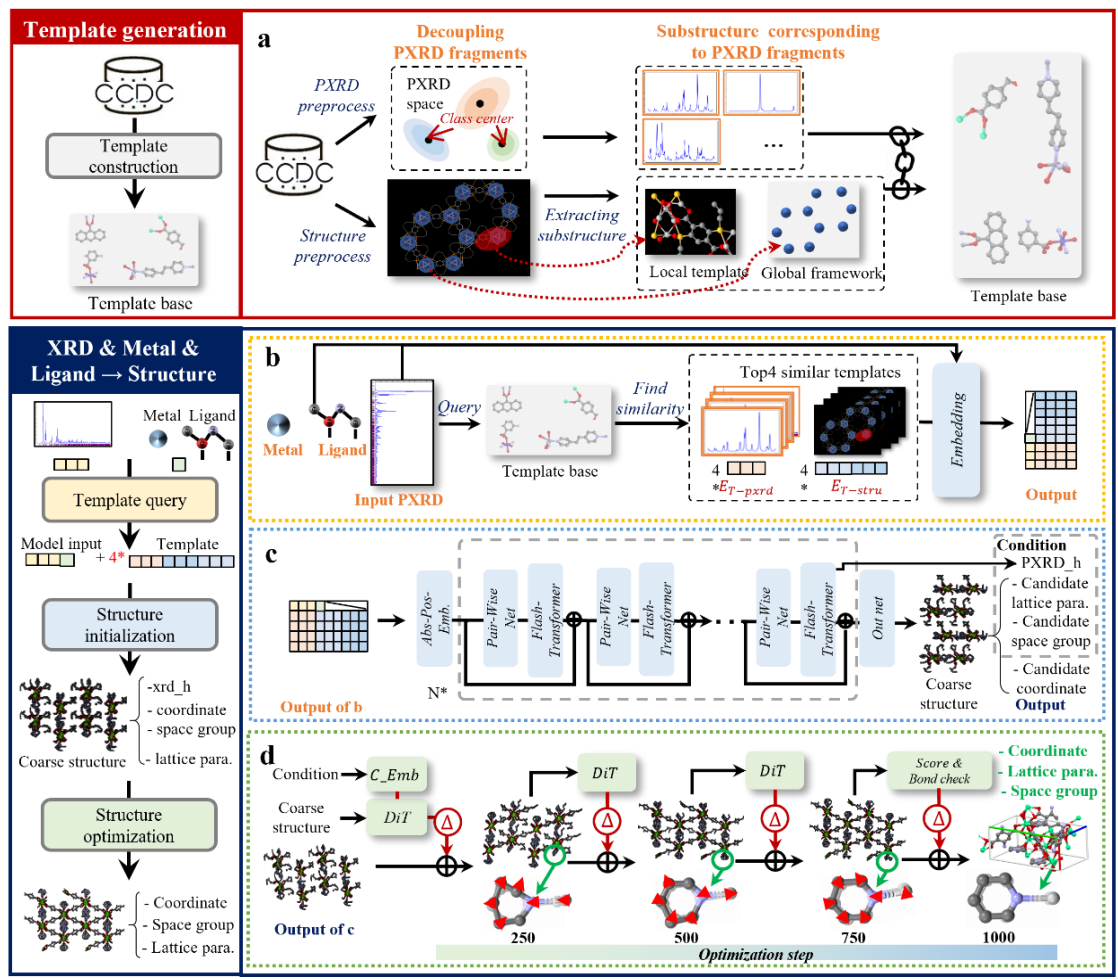

**Fig. 2 | Workflow of USTCunfold. a** Template library construction. PXRD patterns are categorized into 900 distinct groups based on peak intervals, followed by K-means clustering to create representative structural templates. These templates include both detailed local structures and global framework types, each linked with their corresponding PXRD patterns to illustrate the relationship between structure and diffraction accurately. **b** PXRD Embedding Module. This module captures high-dimensional representations by comparing input PXRD patterns with the template library to identify the four most similar templates. Both the input and selected templates are processed in a PXRD embedding sub-module, while structural and ligand templates are embedded concurrently. The data is concatenated into a matrix format, with zeroed-out missing elements ensuring proper alignment for analysis. **c** Structure initialization module. A preliminary coarse structure is generated from the embedded matrix using a stacked residual deep network, which integrates outputs from the template block to capture crystallographic invariances and the flash-transformer to link PXRD data with structural features. The output is processed through three sub-networks to predict crystal space groups, unit-cell parameters, and atomic coordinates, creating a coarse structural model for further refinement. **d** Structure optimization module. This module utilizes multiple layers of the DiT and Score & Bond Check Net to refine coarse structures. It processes unit-cell parameters, space group, atomic coordinates, and PXRD embeddings, retaining structures with scores above 0.9; if none qualify, refinement iterates for improved accuracy in crystal structure predictions.

## End-to-end network architecture and training

The comprehensive workflow of UstcUnfold is depicted in Fig. 2. UstcUnfold represents a sophisticated end-to-end neural network that seamlessly integrates a PXRD-pattern-structure template library with crystal geometry and chemical constraints, enabling precise predictions of crystal structures derived from PXRD data. The principal contributions of UstcUnfold are two-fold: i) We developed a comprehensive pre-processing template library that systematically links PXRD patterns to their corresponding substructures. These spectrum-structure pairs were then employed to enrich the low-dimensional input spectra through a multi-sequence alignment process. ii) We introduced a two-stage neural network, where the first stage utilizes a multi-task learning framework to generate an initial coarse crystal structure based on the input spectrum and matched templates. The second stage applies a diffusion-based residual network[24,25] to refine the structure at the atomic level, incorporating crystal geometry and chemical constraints.

To construct the template library, we assembled and utilized over 1.1 million theoretical structure-PXRD pairs (within the two-theta range of 0–90°) across 225 commonly observed space groups from the CCDC database[10]. This process ultimately yielded a curated set of more than 2,000 templates, each corresponding to specific substructures and their associated PXRD patterns. Fig. 2a illustrates the complete workflow for generating the template library. The integrated PXRD data were segmented into 900 intervals, each spanning 0.1°, and categorized based on the locations of their highest-intensity peaks. For each interval category, we employed k-means clustering[11,12] using data from all remaining intervals (excluding that of the highest peak). From each cluster, we selected the Top-$N$ ($N \in [4, +\infty]$) PXRD patterns closest to the cluster centroid as representative candidate pattern for analyzing and extracting structure templates. The structural templates consist of two primary types: (i) Local detailed templates, which capture common metal-ligand bonding patterns and typical coordination environments; and (ii) Global framework templates, which record the distribution of crystal structures across an entire unit-cell. Notably, template extraction focused on metal atom positions, with organic crystal data treated using benzene rings as pseudo-metal atoms for consistency (see details in Methods). Following the template extraction, we verified structural similarity by comparing atomic classifications and relative coordinates, considering only metals and non-metals. Only structurally similar templates with corresponding PXRD patterns were linked and included in the final template library, ensuring accurate pairings of structure and diffraction data.

Next, we present the backbone architecture of UstcUnfold, which consists of three essential modules: the PXRD embedding module (Fig. 2b), the structure initialization module (Fig. 2c), and the structure optimization module (Fig. 2d). The inputs to UstcUnfold include PXRD data, metal types, and ligand structures, aligning with conventional methods for unfolding PXRD data. Each module serves a distinct function in the overall process of structure determination. The PXRD embedding module is specifically designed to capture the high-dimensional semantic representations of the input data (Fig. 2b). It initiates by comparing the input PXRD patterns against a template library, identifying the top four templates that exhibit the highest similarity to the input PXRD data. These top-four PXRD patterns, in conjunction with the input PXRD data, are subsequently processed through a dedicated PXRD embedding sub-module. Concurrently, both the corresponding structure templates and the input ligand structure undergo embedding via a structure embedding[26] sub-module. The embedded PXRD and structural data are then concatenated and organized into a matrix format, ensuring that each row corresponds either to the input or to a similar template's PXRD and structural information. To maintain alignment within this matrix, any missing elements in rows have been zeroed out. Based on the embedded matrix, the structure initialization module generates a preliminary coarse structure (Fig. 2c). This module employs a stacked residual deep network[13], wherein each layer integrates a template block derived from variated evoformer[14,15] and a flash-transformer net[16]. Prior to stacking, absolute position embedding[17] is introduced into each sequence within the transformer to distinguish between PXRD and structural hidden features. The template block network identifies crystallographic invariances and variations across similar PXRD patterns, while the flash-transformer net captures the relationships between PXRD data and structural information within the same sequence. Through residual stacking of these networks, the model effectively captures both intra- and inter-sequence interactions, resulting in more accurate structural predictions. The final output is subsequently fed into three sub-networks that predict crystal space groups, unit-cell parameters, and atomic coordinates (including types and xyz positions), respectively. Collectively, these predictions define a coarse structural model that also serves as input for the subsequent module alongside the hidden features of the original input PXRD data—this being a direct output of the residual network[27]. The final structure optimization module (Fig. 2d) comprises multiple layers of Diffusion Transformer (DiT)[18] integrated with a chemically constrained network, known as the Score & Bond Check Net. This module accepts as input the unit-cell parameters, space group, atomic coordinates (including types and xyz positions), and PXRD embeddings obtained from the preceding stage. It refines the coarse structure through a diffusion-based process[19] that parallels classical crystal structure refinement[28]. The chemically constrained network is pre-trained to ensure compliance with established chemical rules regarding bond lengths and angles while simultaneously evaluating the congruence between the predicted structure and PXRD data. During the DiT refinement process, only those structures achieving a score exceeding 0.9 (on a scale of 1) are retained. In instances where no structure meets this criterion, the process re-samples the coarse structure and initiates another cycle of refinement. This iterative mechanism of scoring and adjustment enables UstcUnfold to deliver highly accurate crystal structures.

The training of UstcUnfold consists of two distinct phases: unsupervised pre-training of sub-modules and supervised training of the network. The unsupervised pre-training includes the following components: (i) a masked pre-training approach for the flash-transformer, designed to enhance the model's resilience against lower-quality PXRD data, such as datasets characterized by missing peaks or reduced two-theta angles. Specifically, 15% of the input tokens are randomly masked, with unmasked sequences employed to predict these masked tokens[29]; (ii) independent training of the Score & Bond Check Network, which aims to ensure that bond lengths and angles conform to established chemical rules while correlating predicted crystal structures with PXRD data. This process involves generating erroneous structure-PXRD pairs through substitution, deletion, or modification of structural components. These erroneous pairs are then trained alongside authentic structure-PXRD pairs, enabling the network to assign high scores to accurate pairings while penalizing inaccurate ones. Upon the completion of the pre-training phase, we proceed to supervised training, which is further subdivided into two components: (i) independent training of the structure initialization and structure optimization modules. The structure initialization utilizes Mean Absolute Error (MAE) for crystal atomic structures and unit-cell parameters, alongside cross-entropy loss for space groups. This approach leverages a multi-task learning framework[20] to jointly optimize these three outputs. The primary emphasis is placed on atomic coordinates to yield a more accurate coarse structure and facilitate subsequent refinements. In the training of the structure optimization module, the loss associated with atomic coordinates is defined as Root Mean Square Error (RMSE), while losses for unit-cell parameters and space groups adhere to similar criteria as in the previous module, employing Gaussian noise-augmented data for training; (ii) end-to-end training of the entire UstcUnfold model, where the main focus centers on optimizing outputs from the structure optimization module following continuous execution of UstcUnfold.

The training data employed in this study is exclusively derived from publicly available crystal data within CCDC[10], with corresponding theoretical powder X-ray diffraction (PXRD) patterns computed for each crystal structure. Notably, crystal data belonging to the P21/c and P-1 space groups account for over 59% of the total sample size in the CCDC, resulting in a significant imbalance in the distribution of training data[21-23]. Furthermore, atomic distributions across different space groups exhibit considerable variability among crystal structures. To enhance the model's capability to predict structures across all possible space groups for crystalline solids—particularly those with limited sample sizes—we implement a proportional sampling strategy. For space groups containing more than 2,000 samples, we adaptively sample between 1% and 50% of the dataset, inversely proportional to their respective sample sizes. Conversely, for those with fewer than 2,000 samples, a fixed sampling rate of

50% is applied. This methodology facilitates the construction of a comprehensive training and validation dataset while designating unsampled data as the test set.

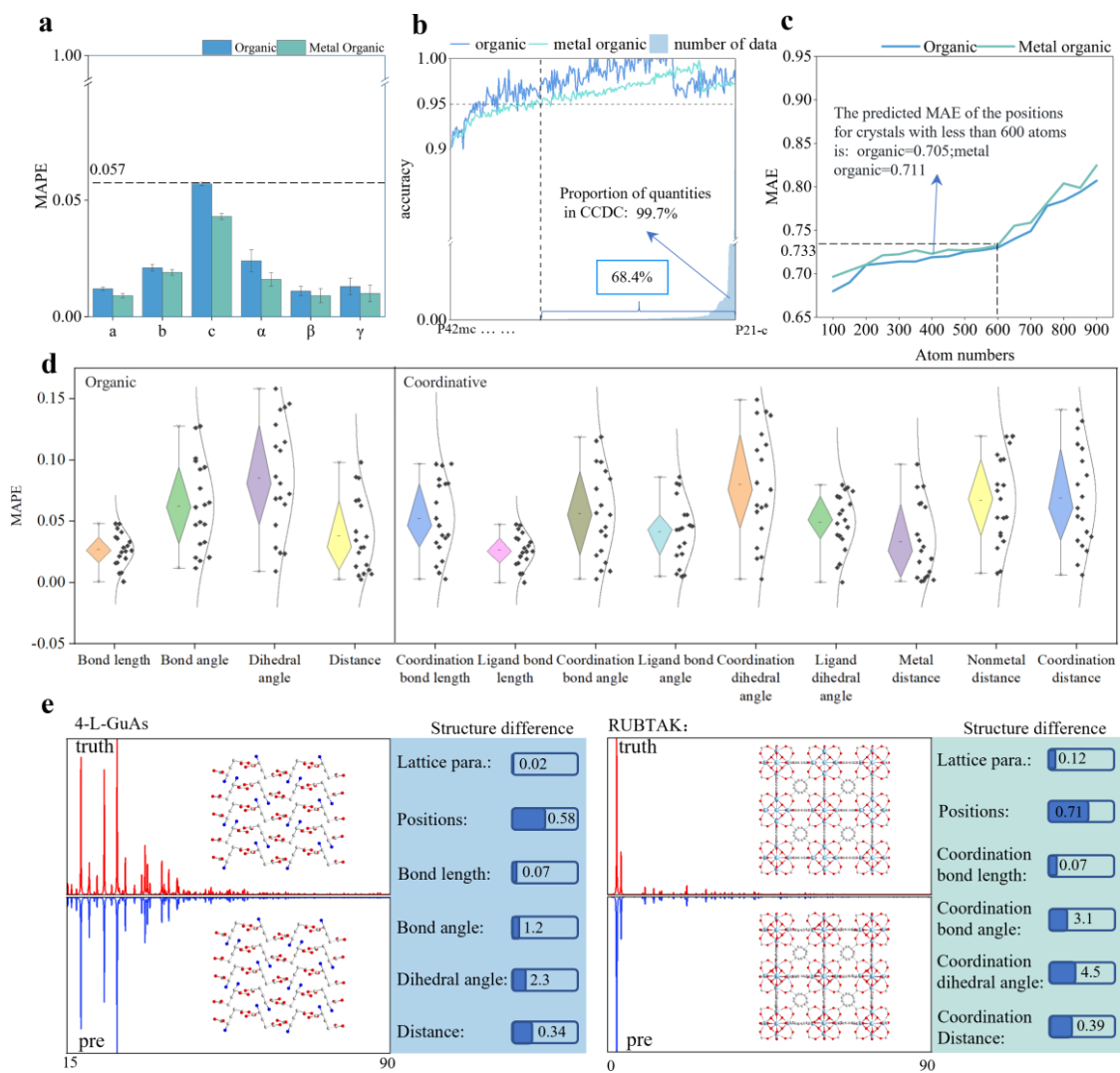

**Fig. 3 | Performance evaluation of UstcUnfold in predicting key crystal structures. a** Prediction accuracy for unit-cell parameters (a, b, c, α, β, and γ), represented by the Mean Absolute Percentage Error (MAPE). UstcUnfold consistently achieved MAPE values below 0.05, with the maximum error limited to 5.7%. **b** Space group classification accuracy across 225 space groups, with over 95% accuracy achieved in 154 groups, encompassing 99.7% of the CCDC dataset. **c** Mean Absolute Error (MAE) for atomic coordinates across 486 samples containing fewer than 600 atoms. Average prediction errors were 0.705 for organic and 0.711 for coordinative compounds, demonstrating minimal performance variance with respect to atom count. **d** MAPE analysis of bond lengths, bond angles, dihedral angles classified into organic/coordinative bonds, and non-bonded atomic distances classified into metal-metal/coordination/non-metal distances, with all metrics achieving MAPE values below 0.1. **e** Two representative examples from both organic and coordinative compounds, showing predicted versus actual crystal structures, alongside corresponding calculated PXRD patterns, highlighting UstcUnfold's high accuracy in structural reconstruction.

## Accuracy across crystalline solids in all possible space groups

UstcUnfold demonstrates commendable accuracy in predicting crystal structures across all possible space groups, utilizing input data that includes PXRD patterns, metal types, and ligand structures. The prediction task is segmented into three fundamental components: unit-cell parameters, space group classification, and atomic coordinates (which encompass atomic types and xyz positions), as previously mentioned. Collectively, these elements delineate the overall spatial distribution, symmetry constraints within the unit-cell, and specific atomic arrangements, thereby facilitating the reconstruction of detailed crystal structures.

In Fig 3, we present the prediction results for these three structural tasks, showcasing examples from both organic and coordination compounds. To rigorously evaluate UstcUnfold's predictive performance, a test set comprising 500 crystal structures was randomly sampled from the CCDC library along with their theoretical PXRD patterns (5–90°). This test set was evenly divided into two categories—organic and coordinative—with 250 samples each. To ensure comprehensive coverage of space groups, we selected two samples from each of the 25 space groups containing more than 2,000 structures and one sample from those with fewer than 2,000 entries; this approach encompassed a total of 225 distinct space groups.

UstcUnfold's performance in predicting unit-cell parameters, assessed through the Mean Absolute Percentage Error (MAPE), yielded notably favorable results. Predictions for six unit-cell parameters (a, b, c, α, β, and γ) consistently achieved MAPE values below 0.05 (Fig. 3a). Although organic compounds generally exhibited slightly higher prediction errors compared to coordinative ones, the maximum MAPE value across all samples was constrained to 0.057, ensuring that the relative error for unit-cell parameters did not exceed 5.7%.

In terms of space group classification, UstcUnfold demonstrated over 90% prediction accuracy across all 225 space groups and surpassed 95% accuracy for approximately 68.4% of these groups (154 groups). These 154 space groups encompass a total of 1,146,660 structures out of the 1,149,525 entries in the CCDC database—resulting in a classification accuracy of 95% for 99.7% of the dataset (Fig. 3b).

To further assess the capability of UstcUnfold in predicting atomic coordinates, we employed a range of evaluation metrics. The Mean Absolute Error (MAE) for atomic coordinates was analyzed across 486 samples containing fewer than 600 atoms. The average prediction errors were found to be 0.705 for organic compounds and 0.711 for coordinative compounds, indicating that the model achieves high accuracy and efficiency for crystals with fewer than 600 atoms. Furthermore, the number of atoms within this range had a negligible impact on prediction performance (Fig. 3c).

Additionally, the predictions made by UstcUnfold regarding bond lengths, bond angles, dihedral angles, and non-bonded atomic distances were evaluated using Mean Absolute Percentage Error (MAPE) metrics (Fig. 3d). Given the critical structural roles that metals play in coordinative compounds, we further refined our evaluation criteria by categorizing bond lengths and angles into coordinative ones and organic ones in ligands, while classifying non-bonded distances into metal-metal distances, coordination distances, and non-metal distances. Across all 13 metrics assessed, UstcUnfold achieved MAPE values below 0.1, demonstrating exceptional accuracy in predicting essential bond and geometric parameters.

Finally, to visually illustrate UstcUnfold's capability to accurately reconstruct crystal structures, we present two randomly selected examples from both organic and coordinative categories (Fig. 3e). These examples display the predicted structures with their actual counterparts and highlight the discrepancies in their calculated PXRD patterns.

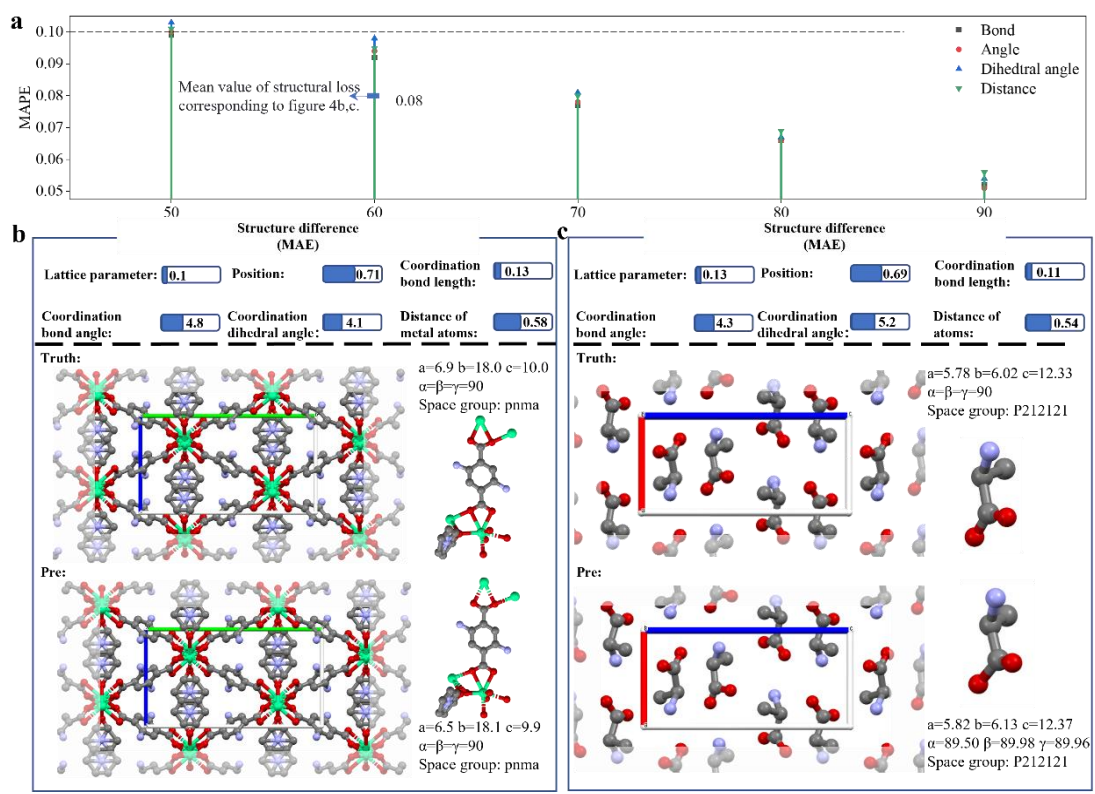

**Fig. 4 | Performance evaluation of UstcUnfold on experimental PXRD data. a** Structural prediction accuracy of UstcUnfold across various 2θ angles, demonstrating maintenance of structural errors (MAPE) below 0.1 at a minimum 2θ of 60°. **b** and **c** Unfolding results of UstcUnfold for both pure organic and organometallic compounds, showcasing accurate space group predictions and a mean absolute error (MAE) of approximately 0.115 for unit-cell parameters. The predicted structures exhibit a high degree of fidelity to the actual structures, as evidenced by the nearly indistinguishable comparison diagrams.

**Experimental pattern examples for predicting crystal structure**

Following meticulous evaluations of UstcUnfold through theoretical PXRD data, we advanced to the exploration of experimental PXRD. It is crucial to acknowledge that classical structure determination from PXRD data presents well-documented challenges, particularly in obtaining high-resolution PXRD datasets that reveal pronounced peaks at elevated 2θ ranges. This undertaking demands sophisticated and costly data-collection methodologies, samples of exceptional purity and crystallinity, and a considerable investment of time. To refine our model, we conducted a series of PXRD pattern analyses across various 2θ ranges to evaluate their influence on the model's predictive performance. Fig. 4a demonstrates that the optimized UstcUnfold maintains structural errors, as measured by MAPE, below 0.1 at a minimum 2θ of 60°, which corresponds to a standard data resolution achievable by most crystalline solids. Additionally, to address the noise and peak-missing issues commonly encountered in experimental PXRD data, we incorporate peak masking of PXRD during the pretraining of UstcUnfold's flash-transformer to enhance the model's robustness to PXRD data of varying precision. Building on this analysis, we further explored the efficacy of UstcUnfold in reconstructing the structures of both organic and coordination compounds from experimental PXRD data collected using standard instrumentation within a brief duration (2-hour scan). All patterns were pre-processed to mitigate the effects of environmental noise. Figures 4b[30] and 4c[31] present the unfolding results, demonstrating that UstcUnfold accurately predicts space groups for both types of compounds, with a mean absolute error (MAE) of approximately 0.115 for unit-cell parameters. The accuracies in predicting atomic coordinates, bond lengths, bond angles of coordination bonds, and dihedral angles are consistent with those derived from theoretical PXRD patterns. Notably, the nearly indistinguishable comparison diagrams between predicted and actual structures for two examples highlight UstcUnfold's capability to effectively resolve complex crystal structures from standard experimental PXRD data.

Therefore, our work not only establishes a rapid and precise AI-driven solution for determining the crystal structures of organic and coordination compounds from widely utilized powder diffraction data—an endeavor of significant importance for natural organic products and biological molecules with coordinating building units—but also has the potential to transform the paradigm that links low-dimensional spectral databases to high-dimensional structural information. Ultimately, this advancement bridges insights from crystal structures to molecular configurations and even electronic structures across various states of matter (solid, liquid, and gas).

**Acknowledgements**

Y.W. acknowledges the National Natural Science Foundation of China (12227901, 62072427), the Project of Stable Support for Youth Team in Basic Research Field, CAS (No.YSBR-005), Academic Leaders Cultivation Program, USTC. P.W. acknowledges the National Natural Science Foundation of China Youth Project (62402472), Natural Science Foundation of Jiangsu Province of China Youth Project (BK20240461). B.Z. and S.W. acknowledge the fund support of the National Natural Science Foundation of China (22071234), the Fundamental Research Funds for the Central Universities (WK9990000113), and the CAS Talent Introduction Program (Category B, KJ9990007009). The authors would like to express their gratitude to Prof. Ya Wang and Prof. Fazhan Shi for their valuable support and helpful discussions.

.